\documentclass[10pt, notitlepage]{article}

\usepackage[english]{babel}

\usepackage{verbatim}
\usepackage{amssymb}
\usepackage{float}
\usepackage{subfig}
\usepackage{graphicx}
\usepackage[T1]{fontenc}
\usepackage[utf8]{inputenc}
\usepackage{authblk}
\usepackage{amsmath,bm}
\usepackage{amsmath,amssymb}
\usepackage{multicol}
\usepackage{amsmath}
\usepackage[left=1.5cm,right=1.5cm,top=2.0cm,bottom=2.0cm]{geometry}
\renewcommand\footnotemark{}

\usepackage[shortlabels]{enumitem}
\usepackage[utf8]{inputenc}

\usepackage{amsmath,amsthm}
\usepackage{cleveref}

\crefname{axiom}{axiom}{axioms}
\Crefname{axiom}{Axiom}{Axioms}

\newtheorem{axiom}{}

\begin{document}
\title{Renormalizing Spacetime}
\author{D.N. Coumbe}
\affil{\small{\emph{The Niels Bohr Institute, Copenhagen University \\Blegdamsvej 17, DK-2100 Copenhagen Ø, Denmark.\\E-mail: daniel.coumbe@nbi.ku.dk}}} 
\date{\small({\today})}
\maketitle

\begin{abstract}

We propose that the consistent field renormalization of gravity requires a specific Weyl transformation of the metric tensor. As a consequence, proper length and time, as well as energy and momentum, become functions of scale. We estimate the functional form of the field renormalization factor by imposing a minimum resolvable distance scale under an infinitesimal Weyl transformation. The derived transformation is applied to two key problems in quantum gravity, its non-conformal scaling, and non-renormalizability.



\vspace{0.25cm}

\end{abstract}

\setlength{\columnsep}{25pt}
\begin{multicols}{2}

\begin{section}{Introduction}

Renormalization is a technique for handling divergent quantities that appear in quantum field theories. An important step in the development of the renormalization procedure was to realize that the parameters appearing in the original theory were actually those defined in the infinite-energy limit, the so-called bare parameters. To make contact with reality, the bare parameters then had to be converted into renormalized parameters defined at a finite energy scale. The bits of the bare parameters that remain following this redefinition can then be reinterpreted as counter-terms, which may act to cancel the divergences. If all infinities can be removed using a finite number of such procedures, the theory is said to be renormalizable. Three of the four fundamental forces have been successfully renormalized, yielding theoretical predictions that agree with experiment to an unprecedented level of accuracy. However, gravity is not renormalizable~\cite{Goroff:1985th}. 

Why does renormalization work so well for the other three forces, but not for gravity? Firstly, consider how renormalization is applied in the case of quantum electrodynamics (QED). The QED Lagrangian is a function of bare charge and mass, as well as bare fields. Renormalizing only the bare parameters of charge and mass will not yield a finite theory. It is crucial that the bare fields are also renormalized. The same is true in perturbative quantum chromodynamics (QCD), where one must renormalize the quark and gluon fields, in addition to the coupling constant and quark mass, in order to obtain a finite theory. In fact, in all quantum field theories, field fluctuations modify bare fields such that they become a function of scale. A bare field $\phi$, for example, is converted into a renormalized field $\tilde{\phi}$ via the process of field renormalization $\tilde{\phi}=\phi Z^{-1/2}(k)$, where $Z(k)$ is a renormalization factor encoding how $\phi$ depends on the so-called coarse-graining scale $k$ (the resolving power of a hypothetical microscope)~\cite{Reuter:2006zq}. However, up until now the renormalization of the gravitational field (the spacetime metric tensor $g_{\mu \nu}$) has been neglected~\cite{Padmanabhan:2015pza}.

A scale-dependent spacetime metric is expected for a number of physical reasons. For example, quantum field theory dictates that the allowed energy of particle-antiparticle pairs in the quantum vacuum increase as a function of the resolution with which spacetime is probed. In combination with general relativity, this implies the metric should fluctuate with a magnitude that depends on scale. Another reason comes from considering the measurement process itself. Operationally, any finite position measurement necessarily requires a scattering particle of non-zero momenta. Thus, gravitational interactions within the observed region are unavoidable. The greater the momentum of the scattering particle, the greater the gravitational perturbation within the region under observation~\cite{Adler:1999bu}. An observer making measurements at a specific coarse-graining scale $k$ will, therefore, deduce a different spacetime metric than if measured at a different scale $k'$. A scale-dependent metric is also predicted by the two leading approaches to quantum gravity, string theory~\cite{Ambjorn:2017vgy} and loop quantum gravity~\cite{Modesto:2008jz}.

Introducing a scale-dependent metric tensor must be done so carefully. For example, scale-dependent vacuum fluctuations are often associated with the vacuum dispersion of light. This is because high-energy photons resolve spacetime on smaller distance scales than their lower energy counterparts, and so are expected to encounter larger metric fluctuations, leading to an energy dependent speed of light. However, recent astronomical data from the \emph{Fermi} space telescope has now definitively shown the speed of light to be independent of energy up to $7.62 E_{P}$ (with a $95 \%CL$) and $\simeq 4.8 E_{P}$ (with a $99 \%CL$)~\cite{Vasileiou:2013vra} for linear dispersion relations, where $E_{P}$ is the Planck energy. We take these experimental results seriously, seeking a scale-dependent metric for which the speed of light is scale-independent. 


Conformal transformations are localized scale transformations for which the angle between vectors is scale-invariant, which includes the angle between null vectors that define the light-cone. Conformal transformations, therefore, seem ideally suited to describing a scale-dependent metric and a scale-independent speed of light. A conformal transformation is a change of coordinates $x^{\mu} \rightarrow \tilde{x}^{\mu}(x^{\mu})$ such that the metric tensor transforms according to $g_{\mu\nu} \rightarrow \Omega^{2}(x^{\mu}) g_{\mu\nu}$, where $\Omega(x^{\mu})$ is a twice differentiable dimensionless function of spacetime coordinates $x^{\mu}$~\cite{Dabrowski:2008kx}. However, a conformal transformation is just a special type of coordinate transformation, a diffeomorphism~\cite{Dabrowski:2008kx}. That is, under a conformal transformation the metric is fundamentally the same, just with shifted coordinates. Thus, a conformal transformation cannot fulfill the requirements necessary for the field renormalization of gravity.

We seek a transformation under which the metric tensor fundamentally changes as a function of resolving power. This motivates the use of the Weyl transformation. A Weyl transformation is similar to a conformal transformation, in that it allows a scale-dependent metric while preserving a scale-independent speed of light, but it is also distinctly different. A Weyl transformation is not a coordinate transformation at all, it is a physical change of the metric tensor $g_{\mu\nu} \rightarrow \Omega^{2}(x^{\mu}) g_{\mu\nu} \equiv \tilde{g}_{\mu\nu}$, where $\tilde{g}_{\mu\nu}$ is a fundamentally different metric that cannot be related to the original via a simple diffeomorphism. 

A normal Weyl transformation allows the metric to change arbitrarily as a function of spacetime coordinates $x^{\mu}$. In this work, we seek the field renormalization of gravity and so we wish to investigate how the spacetime metric $g_{\mu\nu}$ changes as a function of the resolving power $k$. We therefore assume a specific Weyl transformation $g_{\mu\nu}\left(x^{\mu}\right) \rightarrow \Omega^{2}(k) g_{\mu\nu}\left(x^{\mu}\right) \equiv \tilde{g}_{\mu\nu}\left(x^{\mu}\right)_{k}$, where $\Omega(k)$ is an as yet unknown function of the coarse-graining scale $k$~\cite{Reuter:2006zq}. In order that we recover classical physics in the appropriate limit requires that $\Omega(k) \rightarrow 1$ as $k \rightarrow 0$.

For completeness, we therefore state our postulate as:

\begin{axiom}\label{ax1}
At each spacetime point $x^{\mu}$ the bare metric tensor $g_{\mu\nu}\left(x^{\mu}\right)$ is rescaled according to
\begin{equation}
g_{\mu\nu}\left(x^{\mu}\right) \rightarrow \Omega^{2}(k) g_{\mu\nu}\left(x^{\mu}\right) \equiv \tilde{g}_{\mu\nu}\left(x^{\mu}\right)_{k},
\end{equation}

\noindent where $\Omega(k)$ is a dimensionless function of the coarse-graining scale $k$ (the resolving power with which spacetime is probed), subject to the constraint $\Omega(k) \rightarrow 1$ as $k \rightarrow 0$, and $\tilde{g}_{\mu\nu}\left(x^{\mu}\right)_{k}$ is the renormalized $k$-dependent metric tensor. 
\end{axiom}

\noindent  The aim of this paper is to compute the primary physical implications of this Weyl transformation, estimate the functional form of the Weyl factor $\Omega(k)$, and determine whether such a transformation of the metric might be sufficient to make quantum gravity renormalizable. 




\end{section}


\begin{section}{Some immediate implications} 

The length $r$ of any contravariant Riemannian vector $\mathbf{\zeta}$ can be written in tensor form as

\begin{equation}\label{2,1}
r^{2}=g_{\mu\nu}\zeta^{\mu}\zeta^{\nu},
\end{equation}

\noindent where $\zeta^{\mu}$ and $\zeta^{\nu}$ are arbitrary vectors and $g_{\mu\nu}$ is the symmetric metric tensor. Under a Weyl transformation of the metric tensor

\begin{equation}\label{2,2}
g_{\mu\nu}(x^{\mu}) \rightarrow \Omega^{2}(k) g_{\mu\nu}(x^{\mu}) \equiv \tilde{g}_{\mu\nu}(x^{\mu})_{k},
\end{equation}

\noindent the length of any vector $\tilde{r}$ in the new renormalized metric $\tilde{g}_{\mu\nu}(x^{\mu})_{k}$ is then~\cite{Weyl:1918ib}

\begin{equation}\label{2,3}
\tilde{r}^{2}=\tilde{g}_{\mu\nu}(x^{\mu})_{k} \zeta^{\mu}\zeta^{\nu}=\Omega^{2}(k)g_{\mu\nu}\zeta^{\mu}\zeta^{\nu}= \Omega^{2}(k) r^{2}.
\end{equation}

\noindent Likewise, the infinitesimal length of any vector in the renormalized metric is

\begin{equation}\label{2,3.1}
d\tilde{r}^{2}=\tilde{g}_{\mu\nu}(x^{\mu})_{k} d\zeta^{\mu}d\zeta^{\nu}=\Omega^{2}(k)g_{\mu\nu}d\zeta^{\mu}d\zeta^{\nu}= \Omega^{2}(k) dr^{2}.
\end{equation}

\noindent Since the null interval $ds^{2}=0$ and hence speed of light $c=dr/dt$ are invariant under a Weyl transformation, the rescaling of infinitesimal distance of Eq.~(\ref{2,3.1}) immediately implies that time intervals $dt$ must scale according to 

\begin{equation}\label{2,3.2}
dt \rightarrow \Omega(k) dt\equiv d\tilde{t}.
\end{equation}

\noindent Thus, both distance and duration are a function of scale in this scenario, analogous to how distance and duration are a function of relative speed in special relativity. 

The spacetime interval $ds^{2}=g_{\mu\nu}dx^{\mu}dx^{\nu}$ transforms according to 

\begin{equation}\label{2,3.2}
ds^{2} \rightarrow \Omega^{2}(k)ds^{2} \equiv d\tilde{s}^{2},
\end{equation}

\noindent as can be verified using the transformation of Eq.~(\ref{2,2}) on $ds^{2}=g_{\mu\nu}dx^{\mu}dx^{\nu}$. A Weyl transformation therefore changes space-like and time-like intervals, but leaves null intervals invariant. 

A local scale transformation acts on momentum in the opposite way to position~\cite{Qualls:2015qjb}, namely

\begin{equation}\label{qg6}
p \rightarrow \frac{p}{\Omega(k)} \equiv \tilde{p}.
\end{equation}

\noindent The mathematical reason for this is that the product of position and momentum should be dimensionless in natural units. The physical reason is that probing a smaller region of spacetime requires higher frequency modes of momentum~\cite{Qualls:2015qjb}. Energy must then scale identically via

\begin{equation}\label{qg7}
E \rightarrow \frac{E}{\Omega(k)} \equiv \tilde{E},
\end{equation}

\noindent since the speed of light $c=E/p$ must remain invariant under a Weyl transformation. 









Just as quantum fluctuations are present in the QED vacuum, for example, it is reasonable to assume that metric fluctuations should also be present in the gravitational vacuum of flat spacetime~\cite{Padmanabhan:1985jq}. Assuming this is the case then the Weyl transformation of Eq.~(\ref{2,2}) means that the proper length $r$ must also be modified in flat spacetime. 

Let frame $F'$ be an inertial frame of reference in which some length $L'$ is at rest. According to our Weyl transformation of proper length this means that $L' \rightarrow \Omega(k')L'\equiv \tilde{L}'$, where $L'$ is measured at a coarse-graining scale $k'$ by an observer in frame $F'$. Similarly, let $F$ be a frame in which the length $L'$ is moving with a relative velocity $v_{x}$ in the $x$-direction (here we consider only ($1+1$) dimensions for simplicity). In frame $F$ we then have $L \rightarrow \Omega(k)L \equiv \tilde{L}$, when measuring at a coarse-graining scale $k$ in frame $F$. If $\tilde{L}=\tilde{L}'/ \gamma$, where $\gamma=1/\sqrt{1-v_{x}^{2}/c^{2}}$, then

\begin{equation}\label{qg8}                                                                                                                                                                
L=\frac{L'}{\gamma}\frac{\Omega(k')}{\Omega(k)}.
\end{equation}   

\noindent The modified expression for Lorentz contraction of Eq.~(\ref{qg8}) implies a modified Lorentz transformation of the $x$ coordinate of

\begin{equation}\label{qg9}
x'=\gamma (x-v_{x}t)\frac{\Omega(k)}{\Omega(k')},
\end{equation}

\noindent as can be verified by setting $L'\equiv x'_{2}- x'_{1}$ and $t_{2}=t_{1}$ and performing the standard derivation of length contraction, which returns Eq.~(\ref{qg8}). 

Similarly, we can consider a clock at rest in frame $F'$, which via Eq.~(\ref{2,3.2}) implies $\Delta t' \rightarrow \Omega(k')\Delta t'\equiv \Delta \tilde{t}'$. For an inertial frame $F$ relative to which the clock is moving we have $\Delta t \rightarrow \Omega(k)\Delta t\equiv \Delta \tilde{t}$. If $\Delta \tilde{t}=\gamma \Delta \tilde{t}'$, then

\begin{equation}\label{qg10}
\Delta t =\gamma \Delta t' \frac{\Omega(k')}{\Omega(k)}.
\end{equation} 

\noindent The modified time dilation of Eq.~(\ref{qg10}) implies a modified Lorentz transformation of the $t$ coordinate of

\begin{equation}\label{qg11}
t'=\gamma \left(t-\frac{v_{x}x}{c^{2}}\right)\frac{\Omega(k)}{\Omega(k')}.
\end{equation}

Thus, when observations are made at the same coarse-graining scale $k=k'$ in both inertial frames, or in the limit of small $k$ and $k'$, we recover the usual Lorentz transformations, otherwise they are modified. 

\end{section}

                              
\begin{section}{Estimating $\Omega(k)$}\label{motivation2}




An infinitesimal Weyl transformation is given by $\delta g_{\mu\nu}(k)= \epsilon(k)g_{\mu\nu}$, where $\epsilon(k)$ is dimensionless and $|\epsilon(k)| \ll 1$~\cite{Wu:2017epd,Farnsworth:2017tbz}. Therefore,

\begin{equation}\label{2,3.5}
g_{\mu\nu} \rightarrow g_{\mu\nu} + \delta g_{\mu\nu}(k)= g_{\mu\nu} \left(1 + \epsilon(k) \right) \equiv \tilde{g}_{\mu\nu}(x^{\mu})_{k},
\end{equation}

\noindent hence $\Omega^{2}(k)=1+\epsilon(k)$. Under such a transformation vector length scales as

\begin{equation}\label{2,4}
\tilde{r}^{2}=\left(1+ \epsilon(k)\right)r^{2}= \left(r^{2}+ r^{2} \epsilon(k) \right).
\end{equation}

If the perturbation $\delta g_{\mu\nu}$ is sufficiently weak, $\epsilon(k)$ should admit a perturbative series expansion. Since the Weyl factor must be dimensionless we perform a series expansion in the dimensionless combination $lk$, where $l$ is an arbitrary constant with dimensions of length. The product $r^{2} \epsilon(k)$ in Eq.~(\ref{2,4}) then becomes

\begin{equation}\label{2,5}
r^{2} \epsilon(k) = r^{2} \sum\limits_{n=0}^{\infty} a_{n}\left(lk \right)^{n},
\end{equation}

\noindent where $a_{n}$ are coefficients. 

When the physical radius of curvature, as measured in the renormalized metric $\tilde{g}_{\mu\nu}(x^{\mu})_{k}$, is much greater than $1/k$, the WKB approximation of mode functions can be used to show that the coarse-graining scale $k$ is given by $k \simeq \pi / r$, where $r$ is the linear extension of the averaging region~\cite{Reuter:2006zq}. For small perturbations $\delta g_{\mu\nu}$ about flat spacetime we therefore have

\begin{equation}\label{2,6}
\tilde{r}^{2} \simeq r^{2} + r^{2} \sum\limits_{n=0}^{\infty} a_{n}\left(\frac{l}{\pi r} \right)^{n}.
\end{equation}

\noindent The requirement that $\tilde{r}=r$ as $r \rightarrow \infty$ excludes any terms for which $n < 2$ in Eq.~(\ref{2,6}).\interfootnotelinepenalty=10000 \footnote{\scriptsize Note that negative values of $n$ are also excluded by the condition $\tilde{r}=r$ as $r \rightarrow \infty$.} 

We now impose the condition that there is a minimum resolvable distance scale $\tilde{r}=l$, which will further constrain the functional form of $\epsilon(k)$. The motivation for a minimal length is clear: it is a direct consequence of the uncertainty principle, the equivalence principle and a finite and constant speed of light~\cite{Garay:1994en}. Even a straightforward combination of general relativity and quantum field theory leads to the prediction of a minimal length. The uncertainty principle says that measuring ever shorter distances requires concentrating ever larger amounts of energy within ever smaller regions of spacetime. However, this process cannot continue indefinitely. Eventually, the energy density becomes great enough that the region being measured collapses into a black hole, preventing the measurement of any distance smaller than its Schwarzschild radius. Attempting to probe yet shorter distances by adding yet more energy simply results in a bigger black hole~\cite{Scardigli:1999jh,Garay:1994en}. A fundamental length, therefore, seems to be an inevitable feature of quantum gravity~\cite{Hossenfelder:2012jw}. Arguments based on loop quantum gravity~\cite{Hossenfelder:2012jw,Khriplovich:2004fd,Rovelli:1994ge}, string theory~\cite{Gross:1987ar,Veneziano:1989fc} and other approaches to quantum gravity~\cite{Hossenfelder:2012jw} also support this picture.

The requirement that $\tilde{r}=l$ as $r \rightarrow 0$ excludes any terms for which $n > 2$ in Eq.~(\ref{2,6}). Therefore, the only term that reproduces classical physics in the appropriate limit and is consistent with a minimum length scale is $n=2$. Thus, assuming $\epsilon(k)$ admits such a series expansion and $k \simeq \pi /r$, the only Weyl factor for which $\tilde{r}=l$ as $r \rightarrow 0$ and $\tilde{r}=r$ as $r \rightarrow \infty$ is

\begin{equation}\label{2,7}
\Omega(k) \simeq \sqrt{1+\left(lk\right)^{2}}.
\end{equation}

\noindent The functional form of Eq.~(\ref{2,7}) is in good agreement with existing literature. In particular, calculations of quantum conformal fluctuations~\cite{Padmanabhan:1985jq}, quantum gravitational corrections to the propagator~\cite{Padmanabhan:1998yya}, linearized general relativity~\cite{Adler:1999bu}, lattice quantum gravity~\cite{Greensite:1990jm,Coumbe:2015zqa,Coumbe:2015aev} and the Bekenstein-Hawking area law~\cite{Coumbe:2015aev} are all consistent with Eq.~(\ref{2,7}).

\end{section}

\begin{section}{Application to quantum gravity}

In this section, we apply our Weyl transformation to two open problems in quantum gravity; conformal scaling and renormalization. We do not claim that the proposed transformation of spacetime definitively solves these, or any other, problems in quantum gravity. Rather, the hope is that these results capture some key features of a solution that can be further explored and expanded upon.

\begin{subsection}{High-energy conformal scaling}\label{conformal}

There exists a long-standing argument first put forward by Banks~\cite{Banks:2010tj} and later detailed by Shomer~\cite{Shomer:2007vq} that suggests gravity cannot possibly be a renormalizable quantum field theory. Their argument is that since a renormalizable quantum field theory is a perturbation of a conformal field theory (CFT) by relevant operators, the high-energy spectrum of any renormalizable quantum field theory should be equivalent to that of a CFT. However, this is not true for gravity~\cite{Shomer:2007vq}. We now review this problem in more detail before proposing a possible solution.

Using dimensional analysis, the extensivity of the quantities involved and the fact that a finite-temperature conformal field theory has no dimensionful scales other than the temperature $T$, it follows that the entropy $S$ of any $4-$dimensional CFT must scale according to

\begin{equation}\label{cs1}
S \sim a(rT)^{3},
\end{equation}

\noindent and the energy $E$ according to

\begin{equation}\label{cs2}
E \sim b r^{3}T^{4},
\end{equation}

\noindent where $a$ and $b$ are dimensionless numerical coefficients and $r$ is the radius of spacetime under consideration. From Eqs.~(\ref{cs1}) and~(\ref{cs2}) we have

\begin{equation}\label{cs3}
S \sim \frac{a E}{bT}.
\end{equation}

\noindent From Eq.~(\ref{cs2}) we obtain

\begin{equation}\label{cs4}
T \sim \frac{E^{\frac{1}{4}}}{b^{\frac{1}{4}}r^{\frac{3}{4}}}.
\end{equation}

\noindent Substituting Eq.~(\ref{cs4}) into Eq.~(\ref{cs3}) we find that the entropy of a CFT is\interfootnotelinepenalty=10000 \footnote{\scriptsize The scaling of Eq.~(\ref{cs5}) differs from that proposed by Shomer~\cite{Shomer:2007vq}, according to which $S \sim E^{\frac{3}{4}}$ in $4$ dimensions, but agrees with that found by Falls and Litim~\cite{Falls:2012nd}. As pointed out by Falls and Litim the scaling relation $S \sim E^{\frac{3}{4}}$ can only be correct if the radius $r$ is assumed to be constant, an assumption that is unjustified in this context.}

\begin{equation}\label{cs5}
S_{CFT} \sim \frac{a}{b^{\frac{3}{4}}} E^{\frac{3}{4}} r^{\frac{3}{4}}.
\end{equation}


For gravity, however, one expects that the high-energy spectrum will be dominated by black holes~\cite{Banks:2010tj,Shomer:2007vq,Falls:2012nd}. The entropy of a semi-classical black hole is

\begin{equation}\label{cs6}
S_{Grav} = \frac{A}{4G} = \frac{\pi r^{2}}{l_{P}^{2}},
\end{equation}

\noindent where $r$ is the Schwarzschild radius of the black hole and we have used $G=l_{P}^{2}$. It is now clear that Eqs.~(\ref{cs5}) and~(\ref{cs6}) do not agree. Thus, Banks and Shomer conclude that gravity cannot possibly be a renormalizable quantum field theory. This is a serious challenge to quantum gravity that demands a resolution. We now show how this problem may be resolved via the proposed renormalization of spacetime. 



Under our Weyl transformation, distance transforms according to $r \rightarrow \Omega(k)r$. The entropy of Eq.~(\ref{cs6}) then becomes

\begin{equation}\label{cs8}
S_{Grav} \rightarrow \frac{\pi \Omega^{2}(k)r^{2}}{l_{P}^{2}}.
\end{equation}

\noindent We now ask what function $\Omega(k)$ makes the high-energy spectrum of gravity scale in the same way as a conformal field theory? That is, what function is necessary to make $S_{Grav}=S_{CFT}$? To answer this question we form the equation

\begin{equation}\label{cs9}
\frac{\pi \Omega^{2}(k)r^{2}}{l_{P}^{2}} = \frac{a}{b^{\frac{3}{4}}} E^{\frac{3}{4}} r^{\frac{3}{4}},
\end{equation}

\noindent which requires that

\begin{equation}\label{cs10}
\Omega^{2}\left(k \right) \simeq C  \frac{l_{P}^{2}}{r^{2}},
\end{equation}

\noindent where we have assumed $E\simeq \pi/r$ and defined the dimensionless constant $C\equiv a/(\pi^{\frac{1}{4}} b^{\frac{3}{4}})$. 


Since the WKB approximation of mode functions implies $k \simeq \pi / r$~\cite{Reuter:2006zq}, Eq.~(\ref{cs10}) therefore tells us that $\Omega(k)$ must behave like $\Omega(k) \rightarrow \sqrt{C}l_{P}k$ in the high-energy limit.\interfootnotelinepenalty=10000 \footnote{\scriptsize In the presence of strong curvature it is possible that $r$ may significantly deviate from $\pi/k$, although in Ref.~\cite{Reuter:2005bb} it was shown that for $S^{4}$ the relation $r \propto 1/k$ remains valid all the way towards $k \rightarrow \infty$. Furthermore, since the high-energy spectrum for gravity is dominated by black holes, increasing $E$ corresponds to increasing the size of the black hole, thereby lowering the curvature at the horizon. It is possible that the approximation $r \simeq \pi/k$ may then become exact in the infinite-energy limit.} However, in the low-energy limit we must recover $\Omega\left(k \right)=1$, in order to reproduce classical physics and the Lorentz group. A Weyl factor of

\begin{equation}\label{cs12}
\Omega\left(k \right) \simeq \sqrt{1+ C\left(l_{P}k\right)^{2}}
\end{equation}

\noindent satisfies both these requirements, since $\Omega(k) \rightarrow \sqrt{C}l_{P}k$ in the high-energy limit ($1/k \ll l_{P}$) and $\Omega(k) \rightarrow 1$ in the low-energy limit $1/k \gg l_{P}$.\interfootnotelinepenalty=10000 \footnote{\scriptsize We can make a consistency check by substituting $\Omega(k) \simeq \sqrt{C} l_{P}/ r$ back into Eq.~(\ref{cs9}) to obtain $(a\pi^{3/4})/b^{3/4}=(a\pi^{3/4})/b^{3/4}$, where we have used $E=\pi /r$. Thus, $S_{Grav}$ scales in the same way as $S_{CFT}$ in the high-energy limit, in that neither scale (both becoming dimensionless constants).} Therefore, if the Weyl factor is that of Eq.~(\ref{cs12}) then the high-energy spectrum of gravity scales in the same way as a CFT, thereby apparently resolving the Banks~\cite{Banks:2010tj} and Shomer~\cite{Shomer:2007vq} argument. Note, a comparison of Eqs.~(\ref{2,7}) and~(\ref{cs12}) implies that $l \propto l_{P}$, where $l_{P}$ is the Planck length. $\Omega\left(k \right)$ will thus only significantly deviate from unity near the remote Planck scale, hence standard physics will remain safe and sound up to all but the most extreme scales. 


\end{subsection}



\begin{subsection}{Renormalization}


The Green's function for a massless free scalar field $\Phi(x^{\mu})$ in flat Euclidean space~\cite{Padmanabhan:1985jq} is given by 

\begin{equation}\label{qg2}
G(x_{i}^{\mu},x_{j}^{\mu}) \equiv \int \mathcal{D} \Phi(x_{j}^{\mu})\Phi(x_{i}^{\mu}) \exp^{\frac{i}{2}S[\Phi]}=\frac{(4\pi^{2})^{-1}}{\left(x_{j}^{\mu} - x_{i}^{\mu} \right)^{2}}.
\end{equation}

\noindent If the proper distance $r^{2}=\left(x_{j}^{\mu} - x_{i}^{\mu} \right)^{2}$ transforms according to Eq.~(\ref{2,7}), with $k \simeq \pi/r$, then the Green's function is modified to





\begin{equation}\label{qg3}
G(x_{i}^{\mu},x_{j}^{\mu}) \rightarrow \tilde{G}(x_{i}^{\mu},x_{j}^{\mu}) \simeq \frac{(4\pi^{2})^{-1}}{\left(x_{i}^{\mu} - x_{j}^{\mu} \right)^{2}+\pi^{2}l^{2}}.
\end{equation}

\noindent In the limit of zero classical distance $(x_{i}^{\mu} \rightarrow x_{j}^{\mu})$ this yields the finite result

\begin{equation}\label{qg4}
\tilde{G}(x_{i}^{\mu},x_{j}^{\mu})_{x_{i}^{\mu} \rightarrow x_{j}^{\mu}} = \frac{(4\pi^{2})^{-1}}{\pi^{2}l^{2}},
\end{equation}

\noindent where $\tilde{r}=l$ is a minimum renormalized distance scale. A Green's function that remains finite in the limit of zero classical distance is known to remove the ultraviolet divergences present in quantum field theories~\cite{Padmanabhan:1985jq,Hossenfelder:2012jw}. 

We now apply the proposed Weyl transformation to a quantum field theoretic description of gravity. Gravity as a perturbative quantum field theory in $d$-dimensional spacetime dictates that momentum $p$ scales with loop-order $\mathcal{L}$ as

\begin{equation}\label{qg5}
\int{p^{A-[G]\mathcal{L}}} dp,
\end{equation}

\noindent where $[G]$ denotes the canonical mass dimension of the gravitational coupling and $A$ is a process-dependent constant that is independent of $\mathcal{L}$~\cite{Weinberg79}. In 4-dimensional spacetime $[G] =-2$, and so the integral of Eq.~(\ref{qg5}) grows without bound as the loop-order $\mathcal{L}$ increases in the perturbative expansion. Thus, gravity as a perturbative quantum field theory is said to be power-counting non-renormalizable. Here, we show that the integral of Eq.~(\ref{qg5}) becomes finite under a specific transformation of momentum. 

Differentiating $\tilde{p}$ with respect to $p$ via Eqs.~(\ref{qg6}) and~(\ref{2,7}) for a coarse-graining scale $k \simeq p$ gives

\begin{equation}\label{qg9}
d\tilde{p}= \left(\Omega(p)\right)^{-3}dp.
\end{equation}

\noindent Substituting Eqs.~(\ref{qg6}) and~(\ref{qg9}) into Eq.~(\ref{qg5}) we find that a Weyl transformation changes the perturbative expansion of gravity in 4-dimensional spacetime to

\begin{equation}\label{qg100}
\int{p^{A+2\mathcal{L}}} dp \rightarrow \int{\frac{p^{A + 2\mathcal{L}}}{\left(\Omega(p)\right)^{3+ A + 2\mathcal{L}}} dp}.
\end{equation}

\noindent In the infrared limit $\Omega(p) \rightarrow 1$, so we have the same scaling as before the transformation, namely $\sim p^{A + 2\mathcal{L}}$. However, in the ultraviolet limit $\Omega(p)$ behaves like $\Omega(p) \sim p$, so the integrand of Eq.~(\ref{qg100}) scales as $\sim p^{-3}$. 

Figure~\ref{Re1} shows the integrand of Eq.~(\ref{qg100}), denoted by $F(p)$, plotted as a function of the momentum scale $p$ (with $l=A=1$) for $\mathcal{L}=0-5$ loop-orders. As can be seen in Fig.~\ref{Re1}, $F(p)$ increases to a finite maximum and then subsequently decreases to zero in the limit $p \rightarrow \infty$ for all loop-orders. Figure~\ref{Re1} therefore provides quantitative evidence that the scale transformation proposed in this work suppresses divergences usually present in the quantum field theoretic description of gravity, raising the exciting possibility that gravity may become renormalizable in this scenario. 

\begin{figure}[H]
\centering
\includegraphics[width=1.0\linewidth]{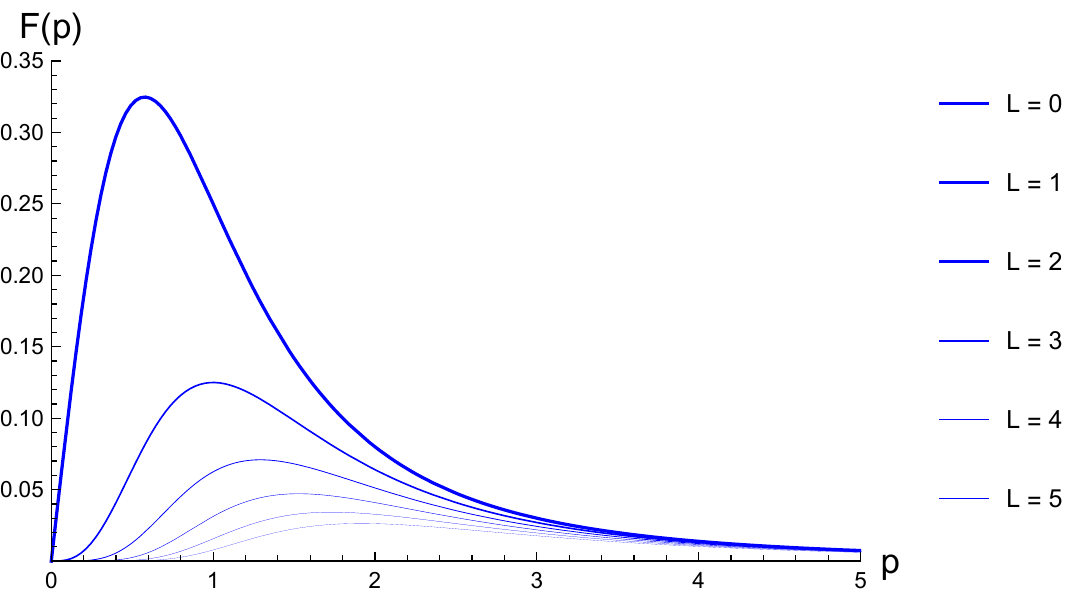}
\caption{\small The integrand of Eq.~(\ref{qg100}) ($F(p)$) plotted as a function of the momentum scale $p$, with $l=A=1$ for $\mathcal{L}=0-5$ loop-orders.}
\label{Re1}
\end{figure}




\end{subsection}

\end{section}


\begin{section}{Discussion and conclusions}

A reduction in the number of spacetime dimensions as a function of scale has now been observed in a wide variety of approaches to quantum gravity~\cite{Ambjorn:2005db,Laihobb,Coumbe:2014noa,Lauscher:2005qz,Horava:2009if,Modesto:2008jz,Atick:1988si}. Almost universally, these approaches find the same dimensional reduction from four at large distances to just two at ultra-small distances. A reduction to two dimensions near the Planck scale would make gravity renormalizable since in this case, the gravitational coupling becomes dimensionless so that Eq.~(\ref{qg5}) is no longer divergent. Furthermore, in two dimensions gravity scales in the same way as a conformal field theory~\cite{Falls:2012nd}, which also resolves the argument outlined in section~\ref{conformal}. However, this so-called dimensional reduction has a number of unphysical consequences including the deformation or violation of Lorentz invariance and typically implies an energy-dependent speed of light~\cite{Coumbe:2015zqa,Coumbe:2015aev}, a prediction that is becoming increasingly constrained by recent astronomical experiments~\cite{Vasileiou:2013vra}. In this work, we have shown that a simple rescaling of the metric tensor can reproduce the desirable features of dimensional reduction, without resorting to a change in dimensionality. 



If the metric tensor runs as a function of the coarse-graining scale $k$ according to $\tilde{g}_{\mu\nu}\left(x^{\mu}\right)_{k}=\Omega^{2}(k) g_{\mu\nu}\left(x^{\mu}\right)$, then proper length and time, as well as energy and momentum, also become running functions of scale. Modified Lorentz transformations follow as a direct consequence of the Weyl transformation considered in this work. We estimate the Weyl factor $\Omega(k)$ by assuming a minimum resolvable distance scale under an infinitesimal Weyl transformation. The application of the derived transformation to two open problems in quantum gravity, the high-energy conformal scaling and renormalization problems, has produced some promising initial results. However, in order to more definitively test whether the proposed transformation of the spacetime metric is sufficient to make quantum gravity renormalizable requires a more detailed study.



\end{section}


\section*{Acknowledgements}

I am especially grateful to Holger Bech Nielsen for extensive discussions on this work. I also wish to thank J. Ambjorn, K. Grosvenor, V. Gueorguiev and C. MacLaurin for their useful comments. I am indebted to the participants of the First Minkowski Meeting on the Foundations of Spacetime Physics for nudging me in this direction. I acknowledge support from the Danish Research Council grant “Quantum Geometry”. 


\bibliographystyle{unsrt}
\bibliography{Master}

\end{multicols}

\end{document}